\documentclass[aps,preprint,preprintnumbers,nofootinbib,showpacs]{revtex4}

\usepackage{graphicx,color,amsmath}
\begin{document}
\title{Collider Phenomenology of Light Strange-beauty Squarks}
\author{Kingman Cheung$^{1)}$, Wei-Shu Hou$^{2)}$}
\affiliation{
$^1$
Department of Physics and NCTS, National Tsing Hua
University, Hsinchu, Taiwan, R.O.C. \\
$^2$
Department of Physics, National Taiwan University, Taipei, Taiwan, R.O.C.}
\date{\today}

\begin{abstract}
Strong mixing between right-handed strange and beauty squarks
is a possible solution to the CP violation discrepancy in $B\to
\phi K_S$ decay as recently suggested by the Belle data. In this
scenario, thanks to the strong mixing one of the strange-beauty
squarks can be as light as 200 GeV, even though the generic
supersymmetry scale is at TeV.
 In this work, we study the production of this light
right-handed strange-beauty squark at hadronic colliders and
discuss the detection in various decay scenarios. Detection
prospect at the Tevatron Run II is good for the strange-beauty
squark mass up to about 300~GeV.
\end{abstract}
\preprint{}
\maketitle

\section{Introduction}

Supersymmetry (SUSY) is the leading candidate for physics beyond the
Standard Model (SM), because it provides a weak scale solution to the
gauge hierarchy problem as well as a dynamical mechanism for the
electroweak symmetry breaking. Usual treatments of SUSY, however,
do not address the problem of flavors. The flavor problem consists
of the existence of fermion generations, their mass and mixing
hierarchies, as well as the existence of CP violation in quark
(and now also in neutrino) mixings, and probably has origins above
the weak scale. In an interesting combination~\cite{Nir} of
Abelian flavor symmetry (AFS) and SUSY, it was pointed
out~\cite{CH,ACH} that a generic feature is the near-maximal $\tilde
   s_R$--$\tilde b_R$ squark mixing. Such a near-maximal mixing allows
for one state to be considerably lighter than the squark mass
scale $\widetilde m$. Such a state, called the strange-beauty
squark $\widetilde{sb}_1$, carries both $s$ and $b$ flavors, and
is bound to impact on $b\to s$ transitions.

It is remarkable that we may have a hint for new physics in CP
violation in $B\to \phi K_S$ decay, which is a $b\to s\bar ss$
transition. The SM predicts that the mixing-dependent CP violation
in this mode, measured in analogous way as the well established CP
violation in $B\to J/\psi K_S$ mode, should yield the same result.
The Belle collaboration, however, has found an opposite sign in
the $B\to \phi K_S$ mode for two consecutive
years~\cite{Belle02,Belle03}. The current discrepancy with SM
prediction stands at a 3.5$\sigma$ level. The result from the
BaBar collaboration in 2003 is at odds~\cite{Browder} with Belle,
but the combined result is still in 2.7$\sigma$ disagreement with
SM expectation. While more data are needed to clarify the
situation, it has been pointed out~\cite{CHNphik} that a light
$\widetilde{sb}_1$ squark provides all the necessary ingredients
to narrow this large discrepancy with SM prediction. It has
(1) a large $s$--$b$ flavor mixing, (2) a (unique) new CP
violating phase, and (3) right-handed dynamics. The latter is
needed for explaining why similar ``wrong-sign" effects are not
observed in the modes such as $B\to K_S\pi^0$ and $\eta^\prime
K_S$. These modes yield consistent results as what was measured in
$B\to J/\psi K_S$. A detailed study of various $B$ decays
suggested~\cite{CHNphik} that $m_{\widetilde{sb}_1} \sim 200$ GeV
and $m_{\tilde g} \sim 500$ GeV are needed, while the squark mass
scale $\widetilde m$ and other SUSY particles can be well above the
TeV scale.

It is clear that a squark as light as 200 GeV is of great interest
since the Tevatron has a chance of seeing it. One should
independently pursue the search for a relatively light
$\widetilde{sb}_1$ squark, even if the $B\to \phi K_S$ CP
violation discrepancy evaporates in the next few years. We note
that a strange-beauty squark, carrying $\sim$ 50\% in strange and
beauty flavor, would lead to a weakening of bounds on beauty
squark search based on $b$-tagging. In this work, we study 
direct strange-beauty squark-pair production, as well as the feed
down from gluino-pair production and the associated production of
$\widetilde{sb}_1$ with a gluino.  It turns out that the dominant
contribution comes from direct squark-pair production as long as
the squark mass is below 300 GeV.  However, for squark mass above
300 GeV, the feed down from gluino-pair production with
$m_{\tilde{g}}=500$ GeV becomes important.  We also study various
decay scenarios of the strange-beauty squarks at the Tevatron,
which is of immediate interest.  The most interesting decay mode
is $\widetilde{sb}_1 \to b/s + \widetilde{\chi}^0_1$, which gives
rise to a final state of multi-$b$ jets plus large missing
energies.  The other scenarios considered are the
$\widetilde{sb}_1$-LSP and the $R$-parity violating 
$\widetilde{sb}_1$ decay possibilities.

The organization of the paper is as follows. In Sec.~II we
recapitulate the features of the model needed for our collider
study. We discuss the production of the strange-beauty squark at
hadronic machines in Sec.~III, and its decay modes and detection
in Sec.~IV.
Conclusion is given in Sec.~V.

\section{Interactions}

We do not go into the details of the model, but mention that the
$d$ flavor is decoupled~\cite{ACH} to evade the most stringent low
energy constraints. The generic class of AFS models~\cite{Nir,CH}
imply a near-maximal $s_R$--$b_R$ mixing, which is extended to the
right-handed squark sector upon invoking SUSY. We focus only on
the $2\times 2$ right-handed strange and beauty squarks, which are
strongly mixed. The mass matrix is given by
\begin{equation}
{\cal L} = - ( \tilde{s}_R^* \; \tilde{b}_R^* )\;
\left( \begin{array}{ll}
   \widetilde{m}_{22}^2 &  \widetilde{m}_{23}^2 e^{-i\sigma} \\
   \widetilde{m}_{23}^2 e^{i\sigma} & \widetilde{m}_{33}^2 \end{array}
  \right ) \;
\left( \begin{array}{c}
      \tilde{s}_R \\
      \tilde{b}_R \end{array} \right ) \;.
\end{equation}
Since the mass matrix is hermitian and the phase freedom has
already been used for quarks, so there remains only one CP violating phase
\cite{CH,ACH,CHNphik}. However, for collider studies it
is not yet relevant. With the transformation
\begin{equation}
\left( \begin{array}{c}
     \tilde{s}_R \\
     \tilde{b}_R \end{array} \right ) = R \,
\left( \begin{array}{c}
     \widetilde{sb}_1 \\
     \widetilde{sb}_2 \end{array} \right ) =
\left( \begin{array}{ll}
    \cos\theta_m& \sin\theta_m\\
    -\sin\theta_me^{i\sigma} & \cos\theta_me^{i\sigma} \end{array} \right ) \;
\left( \begin{array}{c}
     \widetilde{sb}_1 \\
     \widetilde{sb}_2 \end{array} \right ) \;,
\end{equation}
the mass term is diagonalized as
\begin{equation}
{\cal L} = - ( \widetilde{sb}_1^* \; \widetilde{sb}_2^* )\;
\left( \begin{array}{ll}
   \widetilde{m}_{1}^2 &  0 \\
   0 & \widetilde{m}_{2}^2 \end{array}
  \right ) \;
\left( \begin{array}{c}
      \widetilde{sb}_1 \\
      \widetilde{sb}_2 \end{array} \right ) \;.
\end{equation}

\begin{figure}[t!]
\centering
\includegraphics[width=4in]{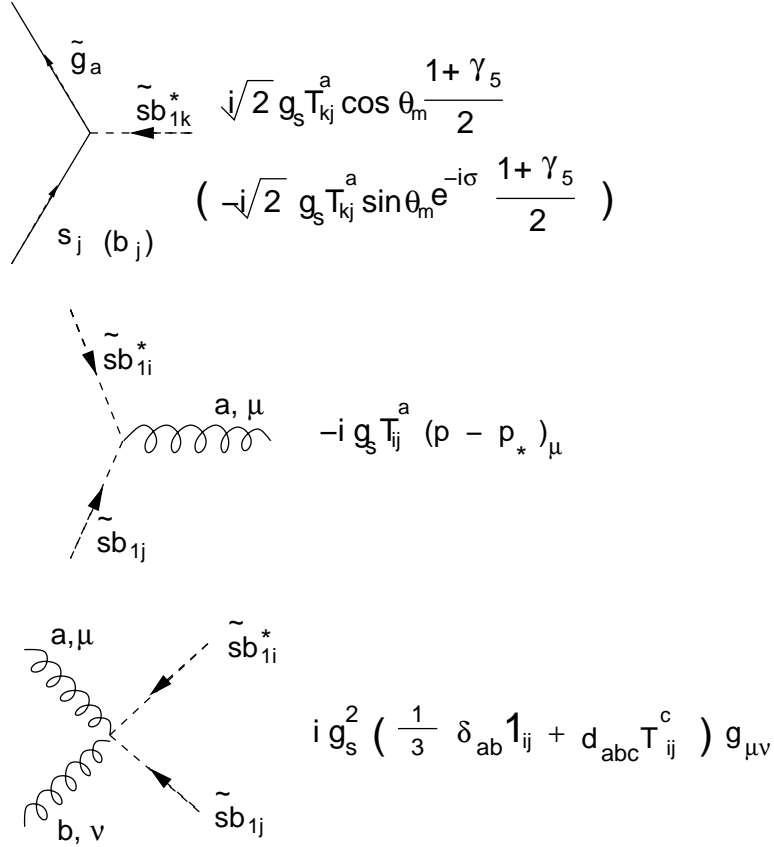}
\caption{\small \label{feyn1}
The relevant Feynman rules used in this work. The momenta are going into
the vertex.}
\end{figure}

The diagonalization matrix $R$ enters the gluino-quark-squark and
squark-squark-gluon interactions. Assuming the quarks are already
in mass eigenbasis, the relevant gluino-quark-squark interaction
in the mass eigenbasis is
\begin{eqnarray}
{\cal L} &=&  -\sqrt{2} g_s T^a_{kj} \left [
 - \overline{\widetilde{g}_a} P_R s_j \widetilde{sb}_{1k}^* \cos\theta_m
 + \overline{\widetilde{g}_a} P_R b_j \widetilde{sb}_{1k}^* \sin\theta_m
   e^{-i\sigma} \nonumber \right .\\
& & \left.
 - \overline{\widetilde{g}_a} P_R s_j \widetilde{sb}_{2k}^* \sin\theta_m
 - \overline{\widetilde{g}_a} P_R b_j \widetilde{sb}_{2k}^* \cos\theta_m
   e^{-i\sigma}
+ {\rm h.c.}
 \right ] \;,\,
 \end{eqnarray}
where $P_R = (1+\gamma^5)/2$, and $a$, $j$, $k$ are the color
indices for gluinos, quarks and squarks, respectively. The
squark-squark-gluon interaction is
\begin{eqnarray}
{\cal L} &=& -i g_s A^a_\mu T^a_{ij} \,\left( \widetilde{sb}_{1i}^*
 {\stackrel{\leftrightarrow}{\partial}}_\mu \widetilde{sb}_{1j}
 + \widetilde{sb}_{2i}^*
 {\stackrel{\leftrightarrow}{\partial}}_\mu \widetilde{sb}_{2j} \right )
 \nonumber \\
&& + g_s^2 (T^a T^b)_{ij} A^{a\mu} A^b_{\mu}
\left( \widetilde{sb}_{1i}^*  \widetilde{sb}_{1j}
 + \widetilde{sb}_{2i}^* \widetilde{sb}_{2j} \right ) \;,
\end{eqnarray}
where
\[
(T^a T^b)_{ij} = \frac{1}{6} \delta_{ab} \delta_{ij} +
\frac{1}{2} (d_{abc} + i f_{abc} ) T^c_{ij} \;.
\]
The relevant Feynman rules are listed in Fig. \ref{feyn1}.

\section{Production at Hadronic Machines} 

We have set the generic SUSY scale at TeV, except for the gluino
and the light strange-beauty squark $\widetilde{sb}_1$, which
could be as light as 500 and 200 GeV, respectively. These masses
are still allowed by the squark-gluino search at the
Tevatron~\cite{cdf-limit}. In fact, these limits are more
forgiving for the present case because $\widetilde{sb}_1$ does not
decay into $b$ quark 100\% of the time.

\begin{figure}[t!]
\centering
\includegraphics[width=5in]{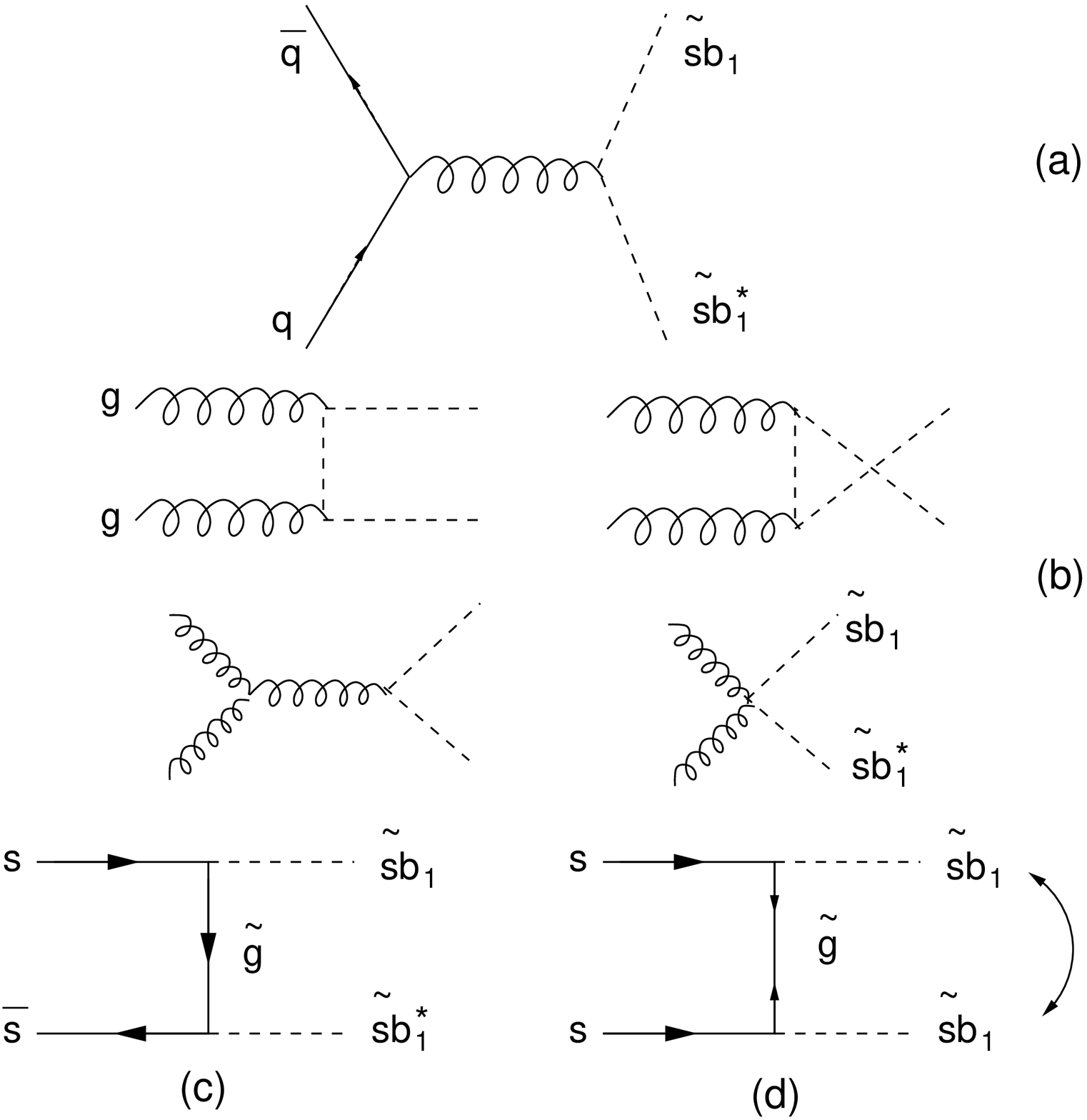}
\caption{\small \label{feyn2} Contributing Feynman diagrams for
(a) $q\bar q \to \widetilde{sb}_1 \widetilde{sb}_1^*$, (b) $g g
\to \widetilde{sb}_1 \widetilde{sb}_1^*$, (c) $s \bar s\ (b \bar
b) \to \widetilde{sb}_1 \widetilde{sb}_1^*$, and (d) $s  s\ (bb)
\to \widetilde{sb}_1 \widetilde{sb}_1$.}
\end{figure}

\subsection{Processes and Formulas}

The production of the strange-beauty squark can proceed via the
following processes.
\begin{enumerate}
\item $q\bar q$ and $gg$ fusion (Fig.~\ref{feyn2}(a) and (b))
\begin{equation}
 q\bar q,\; gg \to \widetilde{sb}_1\,  \widetilde{sb}_1^* \;.
\end{equation}
If the initial state is $s\bar s$ or $b\bar b$, there is an
additional contribution from the $t$-channel gluino exchange
diagram, shown in Fig.~\ref{feyn2}(c). Note that there are also $s\bar b,\,
b\bar s \to \widetilde{sb}_1\,  \widetilde{sb}_1^*$ contributions via
the $t$-channel gluino exchange diagram only.

\item The $ss, bb, \bar s \bar s, \bar b \bar b, sb, \bar
s\bar b$ initial state scattering via $t$- and $u$-channel gluino
exchange diagrams
\begin{equation}
ss,\ sb,\ bb \to \widetilde{sb}_1\,  \widetilde{sb}_1 ,\;\;\; \bar
s \bar s,\ \bar s \bar b,\ \bar b \bar b \to \widetilde{sb}_1^*\,
\widetilde{sb}_1^*\,,
\end{equation}
shown in Fig. \ref{feyn2}(d).
\item
Gluino pair production, followed by gluino decay,
\begin{equation}
q\bar q,gg \to \tilde{g}\tilde{g};\;\;\;
 \tilde{g} \to s \widetilde{sb}_1^*,\; b \widetilde{sb}_1^*,\;
            \bar s \widetilde{sb}_1,\; \bar b \widetilde{sb}_1\,.
\end{equation}
For $s\bar s,\, b\bar b$ in the initial states there are additional
$t$- and $u$-channel diagrams.   Note that $s \bar b, \bar s b \to
 \tilde{g}\tilde{g}$ are also possible through the
$t$- and $u$-channel diagrams.

\item
Associated production of $\widetilde{sb}_1$ with gluino
\begin{equation}
sg,\, bg\, \to \widetilde{sb}_1 \tilde{g} \;,
\end{equation}
followed by gluino decay.

\end{enumerate}

Since the gluino has a mass of at least 500 GeV, we expect the
$t$- or $u$-channel gluino-exchange diagrams to be much smaller
than $q\bar q$ annihilation diagrams.  Moreover, the
$t$- or $u$-channel gluino-exchange diagrams are only relevant for
$s$ or $b$ in the initial state, so the contributions of which
are further suppressed by their parton luminosities.
Nevertheless, we include all those $t$-channel gluino diagrams when the
initial state quarks are $s$ or $b$.
In gluino-pair production we also keep the $t$- and $u$-channel
$\widetilde{sb}_1$-exchange diagrams for the initial state quarks $s$ or $b$.


\vskip0.3cm \noindent \underline{\boldmath\bf Direct production of
$\widetilde{sb}_1 \widetilde{sb}_1^*$} \vskip0.2cm

Let us first introduce some short-hand notation.  The $\hat s,\hat t, \hat u$
are the usual Mandelstem variables.  We define the following
\begin{eqnarray}
\hat t_{\tilde{g}} &=& \hat t - m_{\tilde{g}}^2\,,\;\;\;
\hat u_{\tilde{g}} = \hat u - m_{\tilde{g}}^2\,, \nonumber \\
\hat t_{sb} &=& \hat t - m_{\widetilde{sb}_1}^2\,,\;\;\;
\hat u_{sb} = \hat u - m_{\widetilde{sb}_1}^2\,, \nonumber \\
\beta_{sb} &=&  \sqrt{ 1- \frac{4  m^2_{\widetilde{sb}_1}}{\hat s} }\,, \;\;
\beta_{g} =  \sqrt{ 1- \frac{4  m^2_{\tilde{g}}}{\hat s} }\,, \;\;
\beta_{sbg} = \sqrt{ \left( 1 -
\frac{m^2_{\tilde{g}}}{\hat s} -\frac{m^2_{\widetilde{sb}_1}}{ \hat s}
   \right)^2
 -  4 \frac{m^2_{\tilde{g}}}{\hat s}\, \frac{m^2_{\widetilde{sb}_1}}{\hat s} }
\nonumber \;.
\end{eqnarray}
The subprocess cross section for $q\bar q \to \widetilde{sb}_1
\widetilde{sb}_1^*$ is given by
\begin{equation}
\frac{d\sigma}{d\cos\theta^*}(q\bar q\to \widetilde{sb}_1
\widetilde{sb}_1^*) =
\frac{2\pi \alpha_s^2}{9 \hat s} \beta_{sb} \left[
  \frac{1}{4}( 1- \beta_{sb}^2 \cos^2\theta^*)
  - \frac{m^2_{\widetilde{sb}_1}}{
\hat s} \right ] \;,
\end{equation}
where $\theta^*$ is the central scattering angle in the parton rest frame.
Integrating over the scattering angle $\theta^*$, the cross
section is given by
\begin{equation}
\sigma(q\bar q\to \widetilde{sb}_1 \widetilde{sb}_1^*)=
\frac{2\pi \alpha_s^2}{27 \hat s} \beta_{sb}^3 \;.
\end{equation}
The differential cross section for $gg\to \widetilde{sb}_1
\widetilde{sb}_1^*$ is
\begin{equation}
\frac{d\sigma}{d\cos\theta^*}(gg \to \widetilde{sb}_1 \widetilde{sb}_1^*) =
\frac{\pi \alpha_s^2}{256 \hat s} \beta_{sb} \left ( \frac{64}{3} -
\frac{48 \hat u_{sb} \hat t_{sb}}{\hat s^2} \right ) \left( 1
- \frac{2 \hat s m^2_{\widetilde{sb}_1}}{\hat u_{sb} \hat t_{sb}} +
\frac{2 \hat s^2 m^4_{\widetilde{sb}_1}}
{\hat u_{sb}^2 \hat t_{sb}^2} \right )\;.
\end{equation}
The integrated cross section is given by
\begin{equation}
\sigma(gg \to \widetilde{sb}_1 \widetilde{sb}_1^*)=
\frac{\pi \alpha_s^2}{\hat s} \left[
\beta_{sb} \frac{5 \hat s + 62 m^2_{\widetilde{sb}_1} }{48 \hat s}
 + \frac{ m^2_{\widetilde{sb}_1}}{6 \hat s} \frac{ m^2_{\widetilde{sb}_1}
 + 4 \hat s}{\hat s} \ln \frac{1-\beta_{sb}}{1+\beta_{sb}} \right ] \;.
\end{equation}
For completeness we also give the expressions for $s\bar s, b\bar
b \to \widetilde{sb}_1 \widetilde{sb}_1^*$ cross sections,
\begin{eqnarray}
\frac{d\sigma}{d\cos\theta^*}(s\bar s \to \widetilde{sb}_1 \widetilde{sb}_1^*)
&=& \frac{2 \pi \alpha_s^2 \beta_{sb}}{9 \hat s} \,
\left( \frac{1}{4}( 1- \beta_{sb}^2 \cos^2\theta^*)  -
\frac{m^2_{\widetilde{sb}_1}}{\hat s} \right  ) \nonumber \\
&\times& \left[ 1 - \frac{1}{3} \frac{\hat s}{\hat t_{\tilde{g}} }
\cos^2 \theta_m+ \frac{1}{2} \frac{\hat s^2}{\hat t_{\tilde{g}}^2}
\cos^4 \theta_m\right ] \;. \label{13}
\end{eqnarray}
Integrating over $\cos\theta^*$ gives
\begin{eqnarray}
\sigma(s\bar s &\to& \widetilde{sb}_1 \widetilde{sb}_1^*) =
\frac{2 \pi \alpha_s^2 }{27 \hat s^3} \, \biggr\{ \beta_{sb} \hat s
\left( \hat s \beta_{sb}^2
 - 6 \hat s \cos^4\theta_m + \cos^2\theta_m
 (\hat s + 2 m_-^2 )\right) \nonumber \\
&+&\left. \cos^2\theta_m\left( 2m_-^4
- 3 \hat s \cos^2\theta_m(\hat s + 2m_-^2 )
                               + 2 \hat s m^2_{\tilde{g}} \right )
\log\left(\frac{\hat s + 2 m_-^2 - \beta_{sb} \hat s}
  {\hat s + 2m_-^2 +\beta_{sb} \hat s}
    \right ) \right \} \;, \label{14}
\end{eqnarray}
where $m^2_- = m^2_{\tilde{g}} - m^2_{\widetilde{sb}_1}$.
The cross section for $b\bar b \to \widetilde{sb}_1 \widetilde{sb}_1^*$ can
be obtained by replacing $\cos^2\theta_m\leftrightarrow \sin^2\theta_m$ in
Eqs. (\ref{13}) and (\ref{14}).
On the other hand, the processes $s\bar b,\,
b\bar s \to \widetilde{sb}_1\,  \widetilde{sb}_1^*$ only have the $t$-channel
gluino exchange diagram, and its differential cross section is given by
\begin{equation}
\frac{d\sigma}{d\cos\theta^*}(s\bar b \to \widetilde{sb}_1 \widetilde{sb}_1^*)
= \frac{\pi \alpha_s^2 \beta_{sb}}{9} \,
   \frac{\hat s}{\hat t_{\tilde{g}}^2}
            \cos^2 \theta_m\sin^2 \theta_m
\left( \frac{1}{4}( 1- \beta_{sb}^2 \cos^2\theta^*)  -
\frac{m^2_{\widetilde{sb}_1}}{\hat s} \right  ) \;.
\end{equation}

\vskip0.3cm \noindent \underline{\boldmath\bf Direct production of
$\widetilde{sb}_1 \widetilde{sb}_1$} \vskip0.2cm

Production of $\widetilde{sb}_1 \widetilde{sb}_1$
($\widetilde{sb}_1^* \widetilde{sb}_1^*$) pair requires $ss$, $sb$
or $bb$ ($\bar s \bar s$, $\bar s \bar b$, or $\bar b \bar b$) in
the initial state. The process proceeds via $t$- and $u$-channel
gluino-exchange diagrams, as shown in Fig. \ref{feyn2}(d).  The
differential cross section is given by
\begin{equation}
\frac{d\sigma}{d\cos\theta^*}(s s \to \widetilde{sb}_1
\widetilde{sb}_1) = \frac{\pi \alpha_s^2 \beta_{sb}}{18} \,
\cos^4\theta_m\, m^2_{\tilde{g}} \left[
       \frac{1}{ \hat t_{\tilde{g}}^2 }
     + \frac{1}{ \hat u_{\tilde{g}}^2 } -\frac{2}{3}
       \frac{1}{ \hat t_{\tilde{g}} }
       \frac{1}{ \hat u_{\tilde{g}} } \right ] \;,
\end{equation}
where we have explicitly put in the factor $1/2$ and so $\cos\theta^*$ 
ranges from $-1$ to $1$.
Integrating over the angle the total cross section is
\begin{equation}
\sigma(s s \to \widetilde{sb}_1 \widetilde{sb}_1)
= \frac{\pi \alpha_s^2 \beta_{sb}}{18} \, \cos^4\theta_m\, m^2_{\tilde{g}}
\left[ \frac{4}{ m_-^4 + \hat s m_{\tilde{g}}^2 }
     + \frac{8}{3 \beta_{sb} \hat s}
       \frac{1}{ \hat s + 2 m_-^2 }
       \log \left( \frac{\hat s + 2 m_-^2 - \beta_{sb} \hat s}
                        {\hat s + 2m_-^2 +\beta_{sb} \hat s} \right )
\right]  \;.
\end{equation}
The cross section for $b b \to \widetilde{sb}_1 \widetilde{sb}_1$
can be obtained by replacing $\cos^4\theta_m\leftrightarrow
\sin^4\theta_m$, while that for $s b \to \widetilde{sb}_1
\widetilde{sb}_1$ by replacing $\cos^4\theta_m\leftrightarrow
\cos^2\theta_m \sin^2\theta_m$. Note that, for example, the
amplitude of $b b \to \widetilde{sb}_1 \widetilde{sb}_1$ contains
the phase factor $e^{-2i\sigma}$.  Obviously, when we calculate
the cross section the phase factor drops out.

\vskip0.3cm \noindent \underline{\boldmath\bf Feed down from
gluino-pair production} \vskip0.2cm

We employ a tree-level calculation for gluino-pair production,
though including the next-to-leading order (NLO) corrections \cite{prospino}
the cross section may increase by more than 50\%.  However, the
overall gluino-pair production is small because we have chosen the
gluino mass to be at least 500 GeV.  Whether we include the NLO
correction or not does not affect our conclusion.

Here we give the
tree-level formulas for gluino-pair production, without the squark
in the $t$- and $u$-channels,
\begin{eqnarray}
\frac{d\sigma}{d\cos\theta^*}(q\bar q \to \tilde{g} \tilde{g}) &=&
\frac{2 \pi \alpha_s^2}{3 \hat s} \beta_g \; \frac{ \hat t_{\tilde g}^2 +
  \hat u_{\tilde g}^2 + 2 m_{\tilde{g}}^2 \hat s}{ \hat s^2}, \nonumber \\
\frac{d\sigma}{d\cos\theta^*}(gg \to \tilde{g} \tilde{g}) &=&
\frac{9 \pi \alpha_s^2}{16 \hat s} \beta_g \; \left( 1-
           \frac{ \hat t_{\tilde g} \hat u_{\tilde g}}{\hat s^2} \right )
 \; \left(  \frac{\hat s^2}{ \hat t_{\tilde g} \hat u_{\tilde g}} - 2
          + \frac{4 m_{\tilde{g}}^2 \hat s}{\hat t_{\tilde g}\hat u_{\tilde g}}
   - \frac{4 \hat s^2 m_{\tilde{g}}^4}{ \hat t_{\tilde g}^2 \hat u_{\tilde g}^2
     }  \right ),
\end{eqnarray}
where we have put in the factor of $1/2$ for identical particles
in the final state, and $\cos\theta^*$ is from $-1$ to $1$.  The
integrated cross sections are given by
\begin{eqnarray}
\sigma(q\bar q \to \tilde{g} \tilde{g}) &=&
\frac{8 \pi \alpha_s^2}{9 \hat s} \beta_g \; \left(
 1+ \frac{ 2 m_{\tilde{g}}^2}{\hat s} \right ), \nonumber \\
\sigma(gg \to \tilde{g} \tilde{g}) &=&
\frac{3 \pi \alpha_s^2}{4 \hat s} \; \left[
 -\beta_g \left( 4 + 17 \frac{m_{\tilde{g}}^2}{\hat s} \right )
 + 3 \left( \frac{ 4 m_{\tilde{g}}^4}{\hat s^2} -  \frac{4
  m_{\tilde{g}}^2}{\hat s} - 1 \right ) \, \log \left( \frac{1-\beta_g}
   {1+\beta_g}
\right ) \right ] \;.
\end{eqnarray}

For completeness we also give the cross sections for $s \bar s \to
\tilde{g} \tilde{g}$,
\begin{eqnarray}
\frac{d\sigma}{d\cos\theta^*}(s\bar s \to \tilde{g} \tilde{g}) &=&
\frac{2 \pi \alpha_s^2}{3 \hat s} \beta_g \; \left \{
  \frac{ \hat t_{\tilde g}^2 +
  \hat u_{\tilde g}^2 + 2 m_{\tilde{g}}^2 \hat s}{ \hat s^2}
+ \frac{2}{9} \cos^4\theta_m\left( \frac{ \hat t_{\tilde g}^2}{\hat t_{sb}^2}
   + \frac{\hat u_{\tilde g}^2}{\hat u_{sb}^2} \right )
\right. \nonumber \\
&+& \left. \frac{1}{2} \cos^2 \theta_m\frac{1}{\hat s} \left(
  \frac{\hat s m_{\tilde{g}}^2 + \hat t_{\tilde g}^2}{\hat t_{sb}}
+ \frac{\hat s m_{\tilde{g}}^2 + \hat u_{\tilde g}^2}{\hat u_{sb}} \right)
+ \frac{1}{18} \cos^4 \theta_m\frac{ \hat s m_{\tilde{g}}^2}
   { \hat u_{sb} \hat t_{sb} }  \right \} \;.
\end{eqnarray}
 The formulas for
$b \bar b \to \tilde{g} \tilde{g}$ can be obtained by replacing
$\cos\theta_m$ by $\sin\theta_m$.
Note that $s \bar b, \bar s b \to
\tilde{g}\tilde{g}$ only occur via the $t$- and $u$-channel diagrams, and
the differential cross section is given by
\begin{equation}
\frac{d\sigma}{d\cos\theta^*}(s\bar b \to \tilde{g} \tilde{g}) =
\frac{ \pi \alpha_s^2}{27 \hat s} \beta_g \; \cos^2\theta_m  \sin^2\theta_m
\left \{
 4 \left( \frac{ \hat t_{\tilde g}^2}{\hat t_{sb}^2}
   + \frac{\hat u_{\tilde g}^2}{\hat u_{sb}^2} \right )
+  \frac{ \hat s m_{\tilde{g}}^2}
   { \hat u_{sb} \hat t_{sb} }  \right \} \;.
\end{equation}

We have chosen the mass of gluino to be at least 500 GeV, in order
not to upset lower energy constraints such as $b\to s\gamma$ rate,
and not to violate the bound from direct search at the
Tevatron~\cite{cdf-limit}.  The gluino so produced will decay
into a strange or beauty quark plus the strange-beauty squark
$\widetilde{sb}_1$.  Therefore, gluino-pair production gives two
more jets in the final state than direct production.  Having
more jet activities to tag on may help the detection, especially
when $b$-tagging is employed.  We shall discuss in more detail in
the next section when we treat the decay of the
$\widetilde{sb}_1$. Nevertheless, since the gluino mass is above
500 GeV, the production rate at the Tevatron is rather small.  For
a gluino mass of 500 GeV, the production cross section is 2.9 fb,
which may increase to about 4 fb after taking into account NLO
correction \cite{prospino}.  However, it helps only a little as
far as the strange-beauty squark pair production is concerned,
unless the squark mass is above 300 GeV. We will take this into
account in our analysis.

\vskip0.3cm \noindent \underline{\boldmath\bf Production of
$\widetilde{sb}_1 \tilde{g}$} \vskip0.2cm

There is another process $s(b) g \to \tilde{g} \widetilde{sb}_1$
that can contribute to strange-beauty squark production, but it
requires either $s$ or $b$ in the initial state. The differential
cross section for the process is given by
\begin{eqnarray}
\frac{d\sigma}{d\cos\theta^*}(s g \to \widetilde{sb}_1  \tilde{g}) &=&
\frac{ \pi \alpha_s^2}{192 \hat s} \beta_{sbg}  \cos^2 \theta_m\; \left [
  24 \left( 1- \frac{ 2 \hat s \hat u_{sb}}{ \hat t_{\tilde g}^2}\right )
  -\frac{8}{3}  \right ]   \nonumber \\
&\times& \left[ - \frac{\hat t_{\tilde g}}{\hat s}
   + \frac{2( m_{\tilde{g}}^2 - m_{\widetilde{sb}_1}^2 ) \hat t_{\tilde{g}} }
          {\hat s \hat u_{sb} }\; \left( 1 + \frac{m^2_{\widetilde{sb}_1}}
    { \hat u_{sb}} + \frac{m^2_{\tilde{g}}}{\hat t_{\tilde{g}}} \right )\right]
\;.
\end{eqnarray}
For the $b g$ initial state, the above formula is modified by
changing $\cos^2 \theta_m \leftrightarrow \sin^2\theta_m$.

\begin{figure}[t!]
\centering
\includegraphics[width=5.3in]{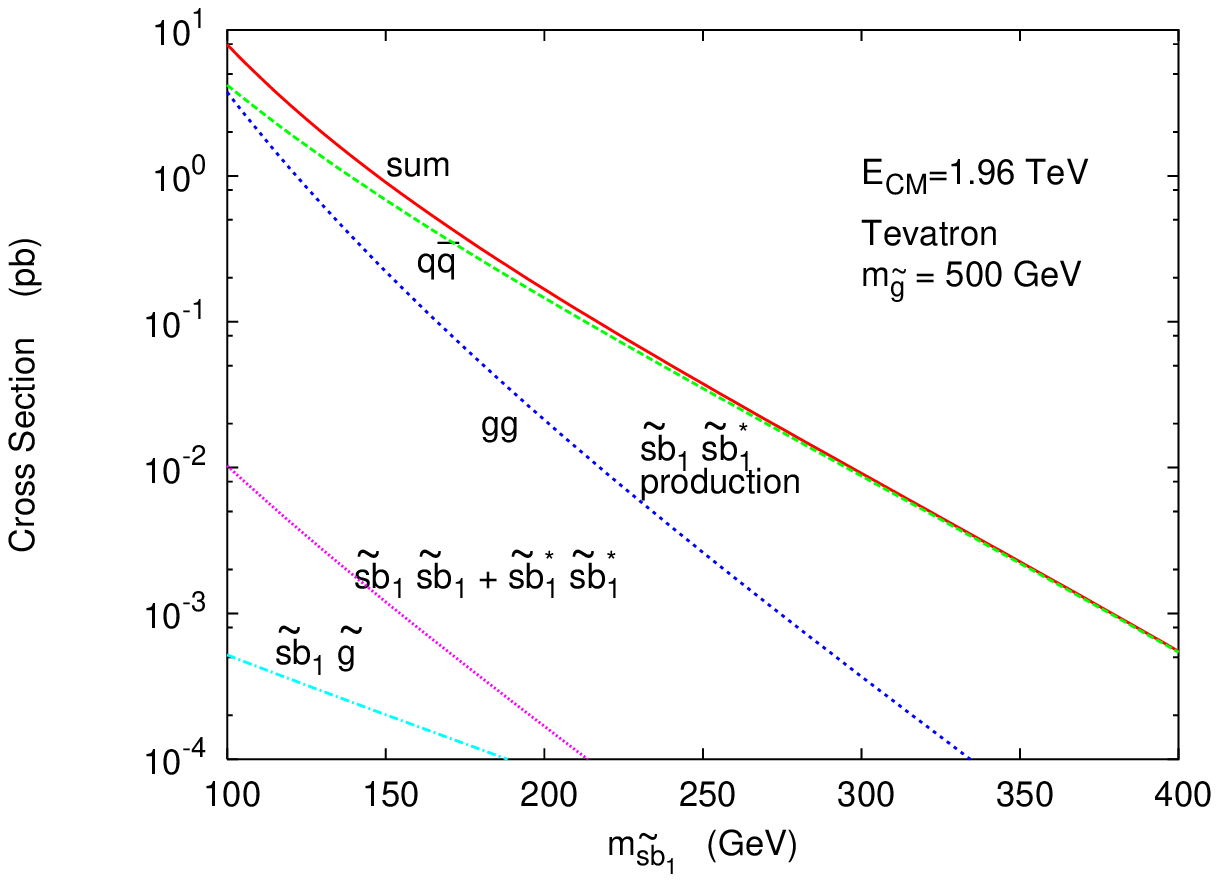}
\includegraphics[width=5.3in]{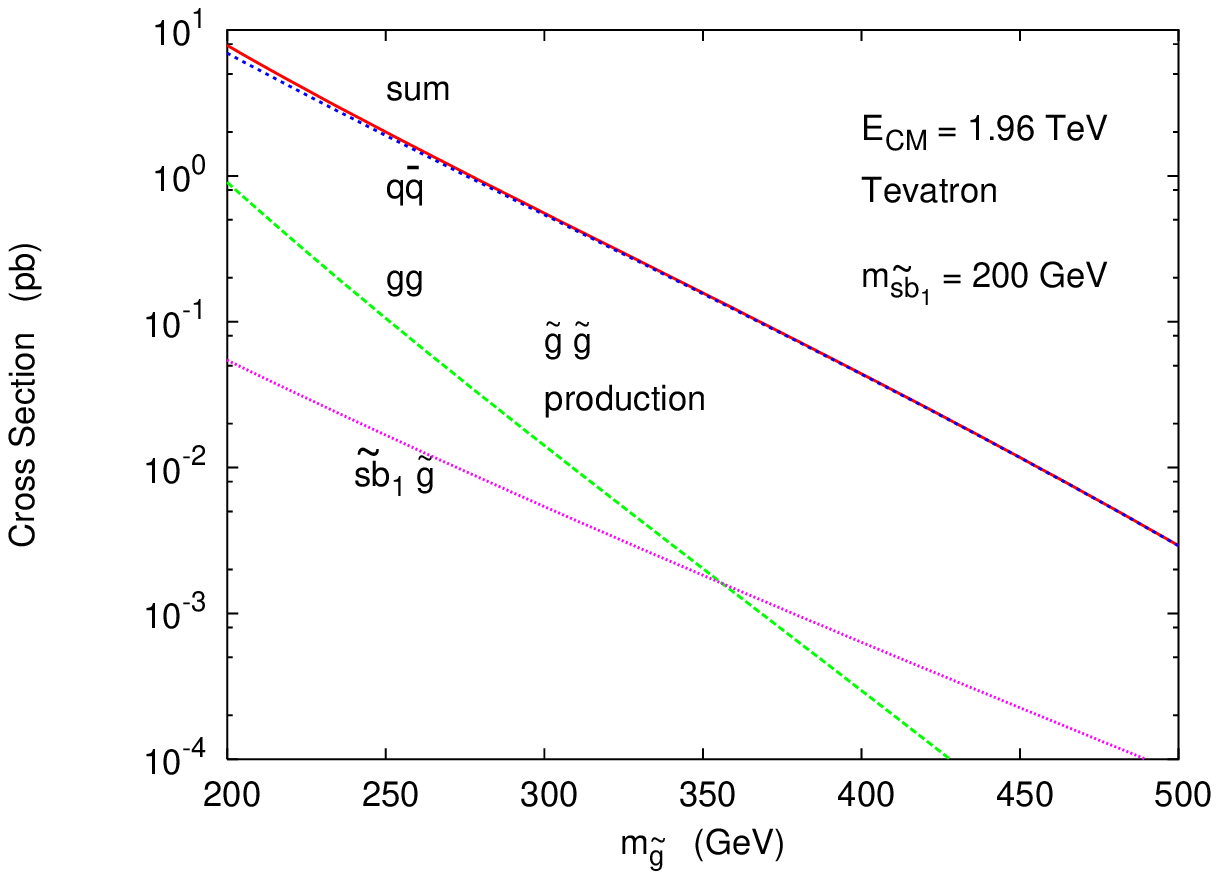}
\caption{\small \label{xs-teva} Total cross section for 
direct production of (a) the $\widetilde{sb}_1 \widetilde{sb}_1^*$ pair
and (b) the $\tilde{g} \tilde{g}$ pair at the Tevatron.
The individual $gg$ fusion and $q\bar q$ annihilation contributions are
shown.
In (a) we also show
$\widetilde{sb}_1 \widetilde{sb}_1 +
              \widetilde{sb}^*_1 \widetilde{sb}_1^*$ production, and
             $\widetilde{sb}_1 \tilde{g} + \widetilde{sb}^*_1 \tilde{g}$
 production, where we have fixed $m_{\tilde{g}}=500$ GeV.
In (b) we also show $\widetilde{sb}_1 \tilde{g} +
\widetilde{sb}^*_1 \tilde{g}$
 production, where we have fixed $m_{\widetilde{sb}_1}=200$ GeV.
}
\end{figure}

\subsection{Production Cross Sections}

\begin{figure}[t!]
\centering
\includegraphics[width=5.3in]{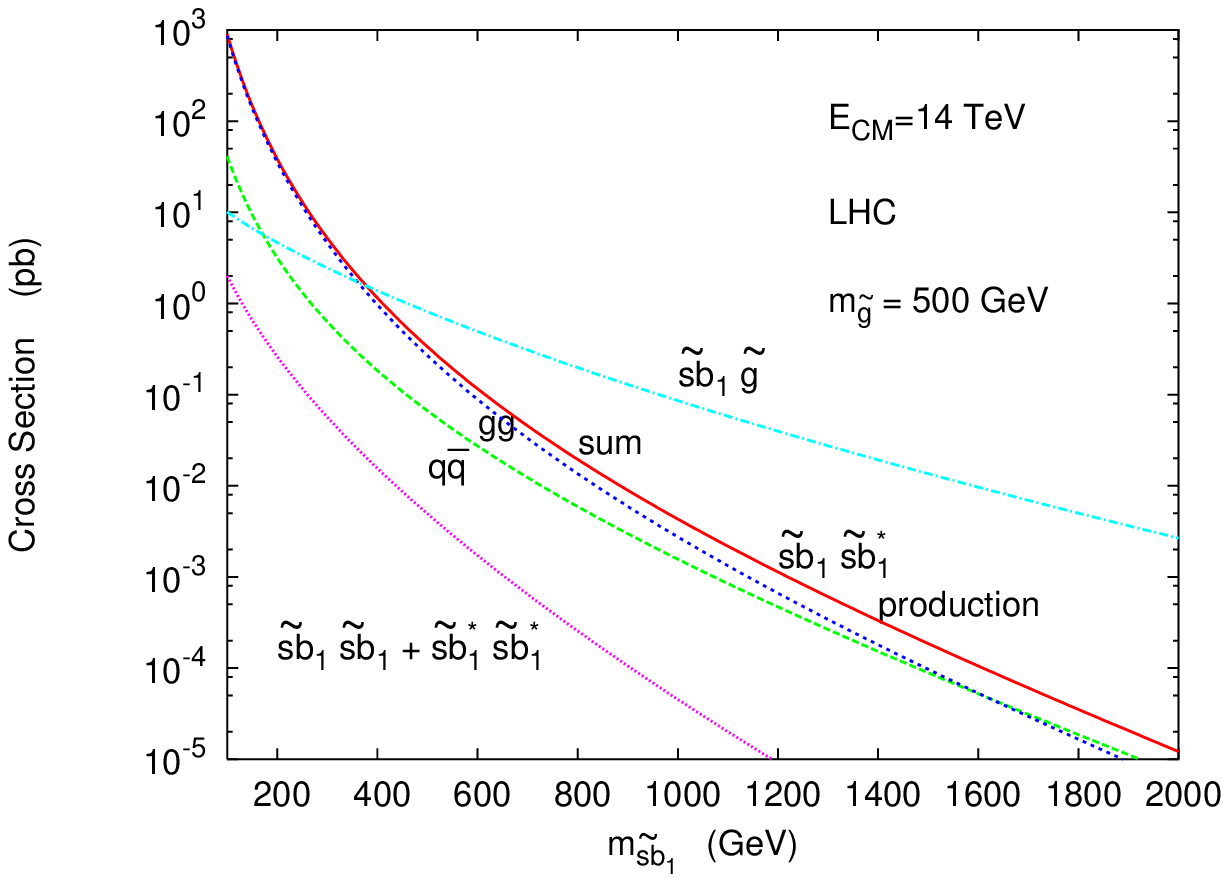}
\includegraphics[width=5.3in]{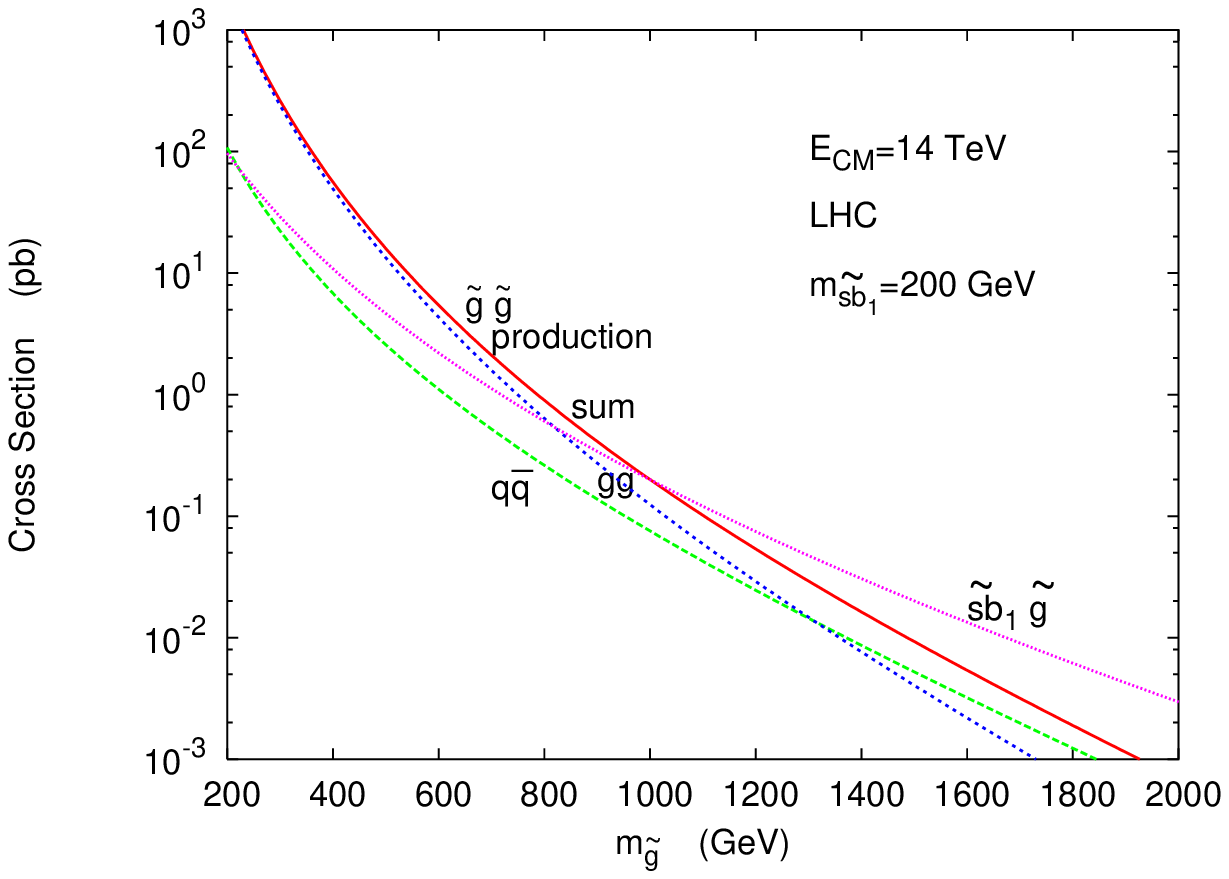}
\caption{\small \label{xs-lhc}
Total cross section for
direct production of (a) the $\widetilde{sb}_1 \widetilde{sb}_1^*$ pair
and (b) the $\tilde{g} \tilde{g}$ pair at the LHC.
The individual $gg$ fusion and $q\bar q$
annihilation contributions are shown.
In (a) we also show
$\widetilde{sb}_1 \widetilde{sb}_1 +
              \widetilde{sb}^*_1 \widetilde{sb}_1^*$ production, and
             $\widetilde{sb}_1 \tilde{g} + \widetilde{sb}^*_1 \tilde{g}$
 production, where we have fixed $m_{\tilde{g}}=500$ GeV.
In (b) we also show  $\widetilde{sb}_1 \tilde{g} +
\widetilde{sb}^*_1 \tilde{g}$
 production, where we have fixed $m_{\widetilde{sb}_1}=200$ GeV.
}
\end{figure}

The cross sections for direct $\widetilde{sb}_1
\widetilde{sb}_1^*$ pair production at the Tevatron are shown in
Fig.~\ref{xs-teva}(a), where we give the individual $gg$ and
$q\bar q$ contributions. As expected, the gluon fusion
contribution is subdominant for $m_{\widetilde{sb}_1} \gtrsim
100$~GeV.
We also show in Fig.~\ref{xs-teva}(a) the same sign
$\widetilde{sb}_1 \widetilde{sb}_1 +
              \widetilde{sb}^*_1 \widetilde{sb}_1^*$ production, and
the associated  $\widetilde{sb}_1 \tilde{g} + \widetilde{sb}^*_1
\tilde{g}$ production.  These processes are three orders of
magnitude smaller than $\widetilde{sb}_1 \widetilde{sb}_1^*$ pair
production, and can be safely ignored at the Tevatron.

Gluino-pair production cross sections at the Tevatron are given in
Fig.~\ref{xs-teva}(b). Similar to squark-pair production,
gluino-pair production is dominated by $q\bar q $ pair
annihilation. For a gluino mass of 500 GeV the cross section is
only a few fb, and thus this contribution becomes comparable to
direct $\widetilde{sb}_1$ pair production only when
$m_{\widetilde{sb}_1} \agt 300$ GeV.  Therefore, at low
$m_{\widetilde{sb}_1}$ the gluino contribution is very small,
while at high $m_{\widetilde{sb}_1}$ the gluino contribution can
extend the sensitivity further.

The situation is different at the LHC.  We show the corresponding
results in Figs. \ref{xs-lhc}(a) and \ref{xs-lhc}(b), respectively.
  We see that
gluon fusion now dominates over $q\bar q$ pair annihilation.
Furthermore, gluino pair production and squark pair production
cross sections are both above 10 pb for
$m_{\widetilde{sb}_1} = 200$ GeV and $m_{\tilde g} = 500$ GeV, and
both contributions have to be taken into account at the LHC.
We also show the associated $\widetilde{sb}_1 \tilde{g} +
\widetilde{sb}^*_1 \tilde{g}$ production cross section in
Figs.~\ref{xs-lhc}(a) and \ref{xs-lhc}(b). These curves are
somewhat misleading, however, that their cross sections become
larger than $\widetilde{sb}_1$-pair (or gluino-pair)
production for large enough $m_{\widetilde{sb}_1}$ ($m_{\tilde
g}$). This is simply because the mass of $m_{\tilde{g}}$ is held
fixed at 500 GeV in Fig.~\ref{xs-lhc}(a) while
$m_{\widetilde{sb}_1}$ is fixed at 200 GeV in
Fig.~\ref{xs-lhc}(b). Therefore, for very large mass the
$\widetilde{sb}_1$- or $\tilde{g}$-pair production become
suppressed.

Before we discuss detection, we need to understand how the
$\widetilde{sb}_1$ squark decays, to which we now turn.

\section{Decay and detection of the strange-beauty squark}

If the SUSY scale is set at TeV, all SUSY particles should be
around this scale, unless one has cancellation mechanisms in the
diagonalization of the neutralino, chargino, or sfermion mass
matrices that allow some of them to become close to the
electroweak scale. The lightness of the $\widetilde{sb}_1$ in our
scenario is a particular example of this type.  This in fact
involves fine-tuning. However, the fine-tuning is
comparable~\cite{ACH} to what is already seen in the quark mixing
matrix. In any case, we do not discuss it further here.

We concentrate on squark-pair production at the Tevatron.  We put
the LHC study aside as its discussion is more intricate, but of
less immediate interest. It is clear from Fig.~\ref{xs-teva}(a)
that the dominant production channels are $gg,q\bar q \to
\widetilde{sb}_1 \widetilde{sb}_1^*$. Gluino-pair production with
$m_{\tilde{g}}=500$ GeV, followed by gluino decay,
is only relevant for $m_{\widetilde{sb}_1}\agt 300$ GeV.
On the other hand, $\widetilde{sb}_1 \widetilde{sb}_1$ and
$\widetilde{sb}_1^* \widetilde{sb}_1^*$ pair production
and the associated production can be safely ignored.

In the following, we take on three situations for the decay of the
strange-beauty squark:

(i) When the $\widetilde{sb}_1$ is the lightest supersymmetric
particle (LSP) and $R$-parity is conserved. This stable
$\widetilde{sb}_1$ case also includes the case when the
$\widetilde{sb}_1$ is stable within the detector but decays
outside.

(ii) The $\widetilde{sb}_1$ is the LSP but $R$-parity is violated
such that it will decay into 2 jets or 1 lepton plus 1 jet.

(iii) The $\widetilde{sb}_1$ is the next-to-lightest
supersymmetric particle (NLSP), and either neutralino (in
supergravity) or gravitino (gauge-mediated) is the LSP such that
$\widetilde{sb}_1$ will decay into a strange or beauty quark plus
the neutralino or gravitino.

Among the three cases we particularly emphasize case (iii), which
is the most popular.  In the SUGRA models, one has
$\widetilde{sb}_1 \to s/b\, \widetilde{\chi}^0_1$, while in
gauge-mediated models $\widetilde{sb}_1 \to s/b\, \widetilde{G}$
or $\widetilde{sb}_1 \to s/b\, \widetilde{\chi}^0_1 \to s/b\;
\gamma \widetilde{G}$. In any case, there will be $b/s$-quark jets
plus a large missing energy in the final state.
We simplify the picture by modelling
the decay as $\widetilde{sb}_1 \to s/b \widetilde{\chi}^0_1$ and
by varying the mass of the neutralino.

\subsection{Stable Strange-beauty Squark}

In this case, the $\widetilde{sb}_1 \widetilde{sb}_1^*$ pair so
produced will hadronize into color-neutral hadrons by combining
with some light quarks. 
Such objects are strongly-interacting massive particles,
electrically either neutral or charged.  If the hadron is
electrically neutral, it will pass through the tracker with little
trace. The interactions in the calorimetry would be rather
intricate, since charge exchange ($\bar d$ replaced by $\bar u$
when passing by a nucleus) can readily occur.
\footnote{An issue arises when the neutral hadron containing the
$\widetilde{sb}_1$ may ``bounce'' into a charged hadron when the
internal $\bar d$ is knocked off and replaced by a $\bar u$, for
example.  The probability of such a scattering depends crucially
on the mass spectrum of the hadrons formed by $\widetilde{sb}_1$.
In reality, we know very little about the spectrum, so we simply
assume a 50\% chance that a $\widetilde{sb}_1$ will hadronize into
a neutral or charged hadron.}
However, the hadron could be electrically charged with equal
probability. In this case, the hadron will undergo ionization
energy loss in the central tracking system, hence behaves like a
``heavy muon''. Let us discuss this possibility since it is more
straightforward.

The energy loss $dE/dx$ due to ionization in the detector material
is very standard \cite{pdg}. Essentially, the penetrating particle
loses energy by exciting the electrons of the material.
Ionization energy loss $dE/dx$
is a function of $\beta \gamma \equiv p/M$ and the charge $Q$ of the
penetrating particle.  The dependence on the
mass $M$ of the penetrating particle comes in through $\beta\gamma$ for
a large mass $M$ and small $\gamma$ \cite{pdg}.
In other words,
 $dE/dx$ is the same for different masses if the $\beta\gamma$ 
values of these particles are the same. For the range of
$\beta\gamma$ between 0.1 and 1 that we are interested in, $dE/dx$
has almost no explicit dependence on the mass $M$ of the
penetrating particle. Therefore, when $dE/dx$ is measured in an
experiment, the $\beta\gamma$ can be deduced, which then gives the
mass of the particle if the momentum $p$ is also measured.  Hence,
$dE/dx$ is a good tool for particle identification for massive
stable charged particles.
In fact, the CDF Collaboration has made a few searches for massive
stable charged particles \cite{cdf1}. The CDF analyses required
that the particle produces a track in the central tracking chamber
and/or the silicon vertex detector, and at the same time
penetrates to the outer muon chamber.
\footnote{In Run II, the requirement to reach the outer muon
chamber may be dropped but it leads to a lower signal-to-background
ratio.
}

The CDF detector has a silicon vertex detector and a central
tracking chamber (which has a slightly better resolution in this
regard), which can measure the energy loss ($dE/dx$) of a particle
via ionization, especially at low $\beta\gamma < 0.85$ ($\beta<
0.65$) where $dE/dx \sim 1/\beta^2$. Once the $dE/dx$ is measured,
the mass $M$ of the particle can be determined if the momentum $p$
is measured simultaneously. Furthermore,  the particle is required
to penetrate through the detector material and make it to the
outer muon chamber, provided that it has an initial $\beta
>0.25-0.45$ depending on the mass of the particle \cite{cdf1}.
Therefore, the CDF requirement on $\beta$ or $\beta\gamma$ is 
(note $\beta\gamma = \beta/\sqrt{1-\beta^2}$)
\[
0.25-0.45 \;\; \alt \;\;\; \beta  \;\;\; < \;\; 0.65  \;\;\;
\Leftrightarrow \;\;\;
0.26-0.50 \;\; \alt \;\;\; \beta \gamma \;\;\; < \;\; 0.86  \;.
\]
The lower limit is to make sure that the penetrating particle can
make it to the outer muon chamber, while the upper limit makes
sure that the ionization loss in the tracking chamber is
sufficient for detection.
CDF has searched for such massive stable charged particles, but
did not find any.  The limits placed on the mass of these
particles are model dependent \cite{cdf-stable}.  Some theoretical
studies on massive stable charged particles exist for gluino LSP
models \cite{gunion}, colored Higgs bosons and Higgsinos
\cite{cho}, and scalar leptons \cite{feng}.

\begin{figure}[t!]
\centering
\includegraphics[width=5in]{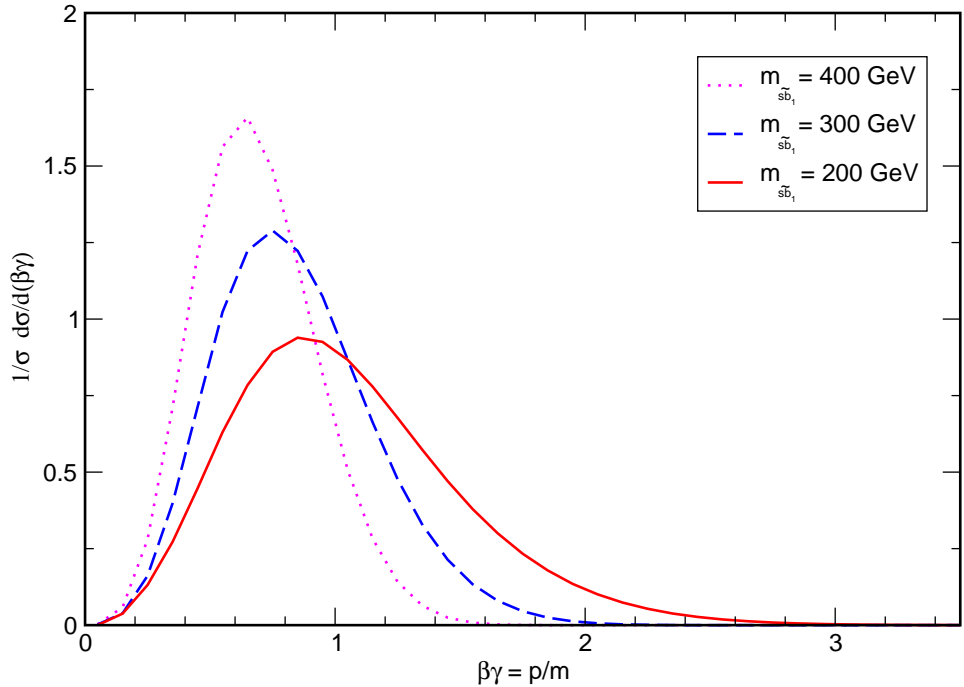}
\caption{\small \label{betagamma} The $\beta\gamma \equiv
p/m_{\widetilde{sb}_1}$ spectrum for squark-pair production at the
Tevatron, where $p$ is the squark momentum.}
\end{figure}

We use a similar analysis for strange-beauty squark pair
production with the squark remaining stable within the detector.
We employ the following acceptance cuts on the squarks
\begin{equation}
\label{cuts}
p_T(\widetilde{sb}_1) > 20 \;{\rm GeV}\,, \qquad |y(\widetilde{sb}_1)|<2.0\,,
\qquad 0.25 < \beta\gamma < 0.85\;.
\end{equation}
In Fig. \ref{betagamma}, we show the $\beta\gamma$ distribution
for direct $\widetilde{sb}_1$ pair production at the Tevatron.  It is
clear that more than half of the cross sections satisfy the
$\beta\gamma$ cut.  This is easy to understand as the squark is
massive such that they are produced close to threshold. In
Table~\ref{stable} we show the cross sections from direct
$\widetilde{sb}_1$ pair production with all the
acceptance cuts in Eq. (\ref{cuts}), for detecting 1 massive
stable charged particle (MCP), 2 MCPs, or at least 1 MCPs in the
final state.   The latter cross section is the simple sum of the
former two.  We have used a probability of 50\%
that the $\widetilde{sb}_1$ will hadronize into a charged hadron.
In the table, we also give the feed down from gluino-pair production
in the parentheses.  It is obvious that the feed down is relatively small
for $m_{\widetilde{sb}_1} \alt 300$ GeV, but becomes significant for
$m_{\widetilde{sb}_1} \agt 300$ GeV.
 Requiring about 10 such events as suggestive
evidence, the sensitivity can reach up to almost
$m_{\widetilde{sb}_1}\simeq300$ GeV with an integrated luminosity 
of 2 fb$^{-1}$.

\begin{table}[t!]
\caption{\small Cross sections for direct strange-beauty squark
pair production at the Tevatron, with the cuts of
Eq.~(\ref{cuts}).
Here $\sigma_{\rm 1MCP}$, $\sigma_{\rm 2MCP}$ denote requiring the
detection of 1, 2 massive stable charged particles (MCP) in the
final state, respectively. Requiring at least one MCP in the final
state corresponds to simply adding the two cross sections.
In parentheses, we give the contribution fed down from direct gluino-pair
production.
\label{stable} }
\medskip
\begin{ruledtabular}
\begin{tabular}{cccc}
 $m_{\widetilde{sb}_1}$ (GeV) & $\sigma_{\rm 1MCP}$ (fb)
  & $\sigma_{\rm 2MCP}$ (fb) & $\sigma_{\geq \rm 1MCP}$ (fb) \\
\hline
$200$ & $41$ $(0.46)$ &   $9.3$ $(0.02)$  & $50$ $(0.48)$  \\
$250$ & $10.9$ $(0.96)$ & $2.8$ $(0.14)$ & $14$ $(1.1)$  \\
$300$ & $3.1 $ $(1.2)$ &  $0.91$ $(0.3)$ & $4.0$ $(1.5)$  \\
$350$ & $0.87$ $(1.3)$&   $0.29$  $(0.43)$ & $1.2$ $(1.8)$  \\
$400$ & $0.23$ $(1.4)$ &  $0.088$ $(0.48)$ & $0.32$ $(1.8)$  \\
$450$ & $0.058$ $(1.4)$ &  $0.024$ $(0.51)$ & $0.082$ $(1.9)$
\end{tabular}
\end{ruledtabular}
\end{table}

\subsection{$\widetilde{sb}_1$ as LSP but $R$-parity is violated}

In this case the $\widetilde{sb}_1$ pair so produced will decay
via the $R$-parity violating terms $\lambda' LQD^c$ or
$\lambda^{''} U^c D^c D^c$ in the superpotential.  In general,
$\lambda'$ and $\lambda''$ couplings are not considered
simultaneously, otherwise it will lead to unwanted baryon decay.
Since nonzero  $\lambda''$ couplings would give only multi-jets in the final
state, which would likely be buried under QCD backgrounds, we only
consider the $\lambda'$ coupling in the following.

By the right-handed nature of the strange-beauty squark in our
scenario, the third index in the $\lambda'$ coupling is either 2
or 3, and we only consider $\lambda'_{ii3},\ \lambda'_{ii2}$ with
$i=1,\ 2$. The strange-beauty squark will decay into $e^-u$ or
$\mu^- c$. Therefore, the strange-beauty squark behaves like
scalar leptoquarks of the first or second generation,
respectively. The decay mode of $\tau^- t$ is not feasible at the
Tevatron. 
The current published limits \cite{lepto-limit} from CDF are 213 GeV
and 202 GeV for the
first and second generation leptoquarks while D\O\ obtained limits of
225 and 200 GeV, respectively.
The latest preliminary limits \cite{lepto-limit} from CDF are 230
and 240 GeV, respectively, while those from D\O\ are 231 and 186
GeV, respectively.  In one of the preliminary plots, the combined
limits from all CDF and D\O\ Run I and II data can push the first
generation leptoquark limit to around 260 GeV, which is very
impressive.~\footnote{
These limits are for the leptoquarks that decay entirely into
charged leptons and quarks. If the leptoquark also decays into a neutrino and
a quark, the corresponding limit is somewhat weaker.}
The sensitivity reach in Run II has been studied in TeV2000 report
\cite{tev2000}.  The reach on the first or second generation
leptoquarks are 235 and 325 GeV with a luminosity of 1 and 10
fb$^{-1}$, respectively.  Apparently, the preliminary limits
obtained by CDF and D\O\ with a luminosity of $\sim 200$ pb$^{-1}$
are already very close to or even surpass the sensitivity reach
quoted in TeV2000 report.  Therefore, we believe that the limit
that can be reached at the end of Run II (2fb$^{-1}$) is very 
likely above 300 GeV.  With an order more luminosity, the limit 
may be able to reach 350 GeV: see the total cross section in 
Fig. \ref{xs-teva}(a)

\subsection{$\widetilde{sb}_1$ is the NLSP}

In this case the $\widetilde{sb}_1$ so produced will decay into a
strange or beauty quark plus the neutralino in the supergravity
framework or the gravitino (or via an intermediate neutralino into
a photon and a gravitino) in the gauge-mediated framework.
Experimentally, the signature is similar, except for the fact that
the neutralino is of order 100 GeV while the gravitino is
virtually massless compared to the collider energy. We simplify
the picture by modelling the decay as $\widetilde{sb}_1 \to s/b
\widetilde{\chi}^0_1$ and by varying the mass of the neutralino.
In addition, we have to check if the strange-beauty squark will
decay {\it within} the detector. In the SUGRA case, the decay rate
is of electroweak strength hence the decay is prompt.  However, in
the gauge-mediated case, the decay rate scales as $\sim 1/F_{\rm
SUSY}^2$, where $\sqrt{F_{\rm SUSY}}$ is the dynamical SUSY
breaking scale.  Therefore, if $\sqrt{F_{\rm SUSY}}$ is so large,
the strange-beauty squark behaves like a {\it stable} particle
inside the detector. Reference~\cite{feng} showed that, for
$\sqrt{F_{\rm SUSY}} \agt 10^7$~GeV the scalar tau NLSP would
behave like a stable particle inside a typical particle detector.
This value applies to the strange-beauty squark as well, up to a
color factor.  If it is stable, one goes back to case (i). So here
we focus on the prompt decay of the strange-beauty squark, which
is considered to be the more popular case.

There are 2 quark jets in the final state of
$\widetilde{sb}_1\widetilde{sb}_1^*$ pair production, each of them
either strange or beauty flavored, and with large missing energy
due to the neutralinos or gravitinos.  We impose the following
cuts on the jets and missing transverse momentum, and we choose
the following $b$-tagging and mistag efficiencies
\footnote{The mistag efficiency is the probability that a non-$b$
jet is detected as a $b$-jet.}
\begin{eqnarray}
p_{Tj} > 15 \;{\rm GeV}\;, &\qquad & |\eta_j|< 2.0\;,
\qquad \not\!{p}_T>40 \;{\rm GeV}, \nonumber \\
\epsilon_{btag} = 0.6  \;, &\qquad & \epsilon_{mis} = 0.05 \;.\nonumber
\end{eqnarray}
Note that the branching ratio of the strange-beauty squark into a
$b$ quark scales as $\sin^2\theta_m$.

We have tested our parton-level Monte Carlo program as follows.
Most events generated pass the jet ($p_{Tj}$ and $|\eta_j|$)
requirements, as long as the mass difference between the
$\widetilde{sb}_1$ and $\widetilde{\chi}^0_1$ is larger than 50
GeV.  Taking $B(\widetilde{sb}_1 \to b \widetilde{\chi}^0_1)=1$ (a
standard $\tilde b$ squark), we verify our input $b$-tagging
efficiency, i.e. the ratio of $0:1:2$ $b$-tagged jets is
$16:48:36$. Choosing the $B(\widetilde{sb}_1 \to b
\widetilde{\chi^0_1})=0.5$ value expected in our scenario, the
ratio of $0:1:2$ $b$-tagged jets becomes $49:42:9$. The double-tag
approach becomes far less effective, but if we only require at
least one $b$-tagged jet in the final state, the overall
efficiency is about $0.5$. On one hand, this is a dilution
compared to the standard $\tilde b$ squark pair production, which
gives overall efficiency of 0.84. On the other hand, the prospect
is still very good for Run II.

In Table~\ref{x-cross}, we show the cross sections in units of fb
for direct squark-pair production, with the squark decaying into either
$s/b$ plus a neutralino at the Tevatron with $\sqrt{s}=1.96$ TeV.
We have set $m_{\tilde{\chi}^0_1}=100$ GeV; other values of
$m_{\tilde{\chi}^0_1}$ do not affect the result in any significant
way, so long as the mass difference between the squark and
neutralino is larger than about 50 GeV.
Note that the production cross section itself is almost
independent of $\sin^2 \theta_m$ and $m_{\tilde{g}}$. This is
because the dominant production channel is the standard QCD
$s$-channel $q\bar q\to \widetilde{sb}_1\widetilde{sb}_1^*$
process and we have imposed $m_{\tilde{g}}\agt 500$ GeV. The
gluino-pair production process, which is also independent of
$\sin^2 \theta_m$, is only $2.9$ fb, and only becomes relevant
for $m_{\widetilde{sb}_1} \agt 300$ GeV.
The branching ratio of the $\widetilde{sb}_1$ into a $b$ quark,
however, scales as $\sin^2 \theta_m$.  We therefore give results
for $\sin^2\theta_m=1,\, 0.75,\, 0.5,\, 0.25$, and for 0, 1, and 2
$b$-tagged jet events. The case for $\sin^2\theta_m=1$ is the same
as a standard $\tilde b$ squark.

\begin{table}[t!]
\caption{\small \label{x-cross} Cross sections in fb for
direct squark-pair production at the Tevatron with $\sqrt{s}=1.96$~TeV,
for 0, 1, 2 $b$-tagged events.  The imposed cuts are $p_{Tj}>15$
GeV, $|\eta_j|<2$, and $\not\!{p}_T>40 \;{\rm GeV}$, 
$b$-tagging efficiency $\epsilon_{btag}=0.6$,
and a mistag probability of $\epsilon_{mis}=0.05$.
In parentheses, we give the contribution fed down from direct
gluino-pair production.}
\medskip
\begin{ruledtabular}
\begin{tabular}{c|ccc|ccc}
 $m_{\widetilde{sb}_1}$ (GeV) & 0 $b$-tag & 1 $b$-tag & 2 $b$-tag
& 0 $b$-tag & 1 $b$-tag & 2 $b$-tag \\
\hline &\multicolumn{3}{c|}{ $\sin^2 \theta_m= 1$} &
\multicolumn{3}{c}{ $\sin^2 \theta_m= 0.75$} \\
$150 $ & $115 (0.11) $   &  $288 (0.54)$ &  $175 (2.2)$  &
 $190 (0.29)$   & $284 (0.89)$  &   $ 104 (1.6)$\\
$200 $ & $26  (0.091) $  &  $70 (0.49)$  &  $47 (2.2)$   &
 $44 (0.27)$    & $70 (0.85) $  &   $ 28 (1.7)$ \\
$250 $ & $6.1 (0.090)$   &  $17 (0.49)$  &  $11 (2.2)$   &
$11 (0.27)$    & $17 (0.85) $  &   $ 6.8 (1.7)$ \\
$300 $ & $1.5 (0.090)$   &  $4.2 (0.49)$ &  $2.9 (2.2)$  &
$2.6 (0.27)$   & $4.2 (0.85)$  &   $ 1.7 (1.7)$ \\
$350 $ & $0.38(0.090)$  &  $1.1 (0.49)$ &  $0.72(2.2)$ &
$0.66 (0.27)$  & $1.1 (0.86)$  &   $ 0.43 (1.7)$ \\
$400 $ & $0.094(0.090)$ &  $0.26(0.49)$&  $0.18 (2.2)$ &
$0.16 (0.27)$  & $0.26 (0.86)$  &  $ 0.11 (1.7)$ \\
$450 $ & $0.022(0.096)$ &  $0.06(0.51)$&  $0.04 (2.2) $ &
$0.038 (0.28)$ & $0.061 (0.87)$ & $ 0.025 (1.7)$ \\
\hline
&\multicolumn{3}{c|}{ $\sin^2 \theta_m= 0.5$} &
\multicolumn{3}{c}{ $\sin^2 \theta_m= 0.25$} \\
$150$ & $283 (0.66)$   & $243 (1.2)  $ & $51 (1.0)$  &
$395 (1.3)$   & $165 (1.1) $ & $17 (0.40)$ \\
$200$ & $68 (0.63) $   & $61 (1.1)   $ & $14 (1.0)$  &
$96 (1.3)$ & $42 (1.1)   $ & $4.6 (0.42)$ \\
$250$ & $16 (0.62) $   & $15 (1.1)   $ & $3.3 (1.0)$ &
$23 (1.3)$& $10 (1.1)   $ & $1.1 (0.42)$ \\
$300$ & $4.0 (0.63)$   & $3.7 (1.1)  $ & $0.84 (1.0)$&
$5.8 (1.3)$& $2.5 (1.1)   $ & $0.28 (0.42)$ \\
$350$ & $1.0 (0.63)$   & $0.93 (1.1) $ & $0.21 (1.0)$&
$1.4 (1.3)$& $0.64 (1.1) $ & $0.071 (0.43)$ \\
$400$ & $0.25 (0.63)$   & $0.23 (1.2) $ & $0.052 (1.1)$&
$0.35 (1.3)$& $0.16 (1.1) $ & $0.017 (0.43)$ \\
$450$ & $0.058 (0.64)$ & $0.053 (1.2)$ & $0.012 (1.0) $&
$0.083 (1.3)$ & $0.037 (1.1)$ & $0.004 (0.42)$
\end{tabular}
\end{ruledtabular}
\end{table}

We see that, for $\sin^2\theta_m\agt 0.5$, if we only require at
least one $b$-tagged jet rather than demanding double-tag, the
cross section does not change drastically as $\sin^2\theta_m$
decreases from 1 to 0.5.
Requiring a minimum of 10 signal events as suggestive evidence for
such a squark, with an integrated luminosity of 2 fb$^{-1}$ the
sensitivity is around 300 GeV, if $\sin^2\theta_m\agt 0.5$. If the
integrated luminosity can go up to 20 fb$^{-1}$, then the
sensitivity increases to 350 GeV.

We emphasize that the double-tag vs single-tag ratio contains
information on $\sin^2\theta_m$, while their sum, when compared with
the standard $\tilde b$ squark pair production, provides
additional consistency check on  cross section vs mass. Such work
would depend on more detailed knowledge of the detector, which we
leave to the experimental groups.

\section{Discussion and Conclusion}

In this work, we have considered the SUSY scenario that the only
light degrees of freedom are the right-handed strange-beauty
squark ($m_{\widetilde{sb}_1}\agt 200$ GeV) and gluino
($m_{\tilde{g}}=500$ GeV). Such a light squark is a result of a
near-maximal mixing in the 2-3 sector of the right-handed squarks,
which is in turn a result of approximate Abelian flavor symmetry.

We have performed calculations for direct strange-beauty
squark-pair production, as well as the feed down from gluino-pair
production and the associated production of $\widetilde{sb}_1$
with gluino. It turns out that the dominant contribution comes
from direct squark-pair production as long as the squark mass is
below 300 GeV. 
As one has to require $m_{\tilde g} \gtrsim 500$ GeV, which comes from
low energy bounds, gluino pair production is in general subdominant.
However, as the squark mass is above 300 GeV, the feed down from
gluino-pair production with $m_{\tilde{g}}=500$ GeV becomes sizable.
Furthermore, many new avenues such as $s\bar s$ ($ss$) $\to
\widetilde{sb}_1\widetilde{sb}_1^{(*)}$ opens up. These are,
however, very suppressed at Tevatron energies because of heavy
gluino mass.

We have studied three decay scenarios of the strange-beauty
squarks that are relevant for the search at the Tevatron, which is
of immediate interest because it can be readily done in the near
future. 
The three decay modes that we have considered are (i)
(quasi-)stable $\widetilde{sb}_1$ as in $\widetilde{sb}_1$-LSP
SUSY or in gauge-mediated SUSY breaking with a very large
$\sqrt{F}$, (ii) $R$-parity violating decay of $\widetilde{sb}_1$
(hence $\widetilde{sb}_1$ behaves like a leptoquark), and (iii)
the popular case of $\widetilde{sb}_1 \to s/b\,\tilde{\chi}^0_1$ decay,
where $\tilde{\chi}^0_1$ is the LSP. 
In the first case, the $\widetilde{sb}_1$ once produced would
hadronize into a massive stable charged particle like a ``heavy
muon", which would ionize and form a track in the central tracking
system and in the outer muon chamber.  This is a very clean
signature.  The sensitivity for Run II with an integrated
luminosity of 2 fb$^{-1}$ is up to about 300 GeV, which may
increase to about 350 GeV with an order more luminosity. 
In the second case, the $\widetilde{sb}_1$ decays like a
leptoquark of the first or second generation.  The best current
limit is 260 GeV (preliminary \cite{lepto-limit}) for the first
generation.  This is already at the sensitivity level of the
Tev2000 study \cite{tev2000} for 2 fb$^{-1}$. With an order more
luminosity, the limit should reach 350 GeV. 
In the last case, $\widetilde{sb}_1 \to s/b\,\tilde{\chi}^0_1$ decay leads
to multiple $b$-jets plus large missing energy in the final state.
The number of $b$-tag events depends on the mixing angle
$\sin\theta_m$, because the branching ratio of $\widetilde{sb}_1
\to b\,\tilde{\chi}^0_1$ scales as $\sin^2\theta_m$.  As long as
$\sin^2\theta_m \agt 0.5$, the sensitivity at the Run II with 2
fb$^{-1}$ goes up to about 300 GeV. 
With improved $b$-tagging in Run II, one can also make use of the
single versus double $b$-tag ratio as well as the $b$-tagged cross
section to determine $m_{\widetilde{sb}_1}$ and the mixing angle
$\sin^2\theta_m$.

At the LHC  $\widetilde{sb}_1\widetilde{sb}_1^*$ and $\tilde g \tilde
g$ pair production cross sections are comparable, with $gg$ fusion
being the dominant mechanism.
Unlike at the Tevatron, the associated production of
$sg\to \widetilde{sb}_1\tilde g$ becomes interesting at the LHC.
Nevertheless, $\widetilde{sb}_1\widetilde{sb}_1$ or
$\widetilde{sb}_1^*\widetilde{sb}_1^*$ pair production remains
relatively unimportant.
With $\widetilde{sb}_1$ as light as 200 GeV,
$\widetilde{sb}_1\widetilde{sb}_1^*$ pair production may be
relatively forward.
On the other hand, $\tilde g\tilde g$ events, followed by $\tilde
g\to \widetilde{sb}_1 \bar s/\bar b$, 
would have extra hard jets to provide more handles.
The $\widetilde{sb}_1\tilde g$ final state, if it can be
separated, can probe the mixing angle $\cos^2\theta_m$ in the
production cross section.
Discovery of the strange-beauty squark at the LHC should be no
problem at all, but the richness demands a more dedicated study,
which we leave for future work.

In conclusion, the recent possible CP violation discrepancy in
$B\to \phi K_S$ decay suggests the possibility of a light
strange-beauty squark $\widetilde{sb}_1$ that carries both strange
and beauty flavors. Such an unusual squark can be searched for at
the Tevatron Run II, with the precaution that $\widetilde{sb}_1$
can decay into a beauty or strange quark, and the standard
$\tilde b$ search should be broadened. Discovery up to 300 GeV is
not a problem, and anomalous behavior in both production cross
sections and the single versus double tag ratio may provide
confirming evidence for the strange-beauty squark.

\section*{Acknowledgments}

This research was supported in part by
the National Science Council of Taiwan R.O.C. under grant no.
NSC 92-2112-M-007-053- and NSC-92-2112-M-002-024, and by 
the MOE CosPA project.

\end{document}